\documentclass[%
 reprint,
superscriptaddress,
nofootinbib,
 amsmath,amssymb,
 aps,
]{revtex4-2}
\usepackage{graphicx}
\usepackage{dcolumn}
\usepackage{bm}
\usepackage[version=4]{mhchem} 
\usepackage{subfig}
\usepackage{float}
\usepackage{caption}
\usepackage{multirow}
\usepackage[colorlinks,
            linkcolor=red,
            anchorcolor=blue,
            citecolor=blue
           ]{hyperref}

\graphicspath{ {figure/} }
\begin{document}

\title{Nonlinear diffusive shock acceleration with upstream escape reproduces DAMPE observations}

\newcommand{\affiI}{%
 State Key Laboratory of Particle Astrophysics, Institute of High Energy Physics, Chinese Academy of Sciences, Beijing 100049, China%
}

\newcommand{\affiII}{%
 Key Laboratory of cosmic ray, Ministry of education, Tibet University, Lhasa 850000, China%
}

\newcommand{\affiIII}{%
 University of Chinese Academy of Sciences, Beijing 100049, China%
}

\author{Han-Xiang Hu}
\email{huhx@ihep.ac.cn}
\affiliation{\affiI}
\affiliation{\affiII}

\author{Xing-Jian Lv}
\email{lvxj@ihep.ac.cn}
\affiliation{\affiI}
\affiliation{\affiIII}

\author{Xiao-Jun Bi}
\email{bixj@ihep.ac.cn}
\affiliation{\affiI}
\affiliation{\affiIII}

\author{Tian-Lu Chen}
\email{chentl@ihep.ac.cn}
\affiliation{\affiII}

\author{Kun Fang}
\email{fangkun@ihep.ac.cn}
\affiliation{\affiI}

\author{Peng-Fei Yin}
\email{yinpf@ihep.ac.cn}
\affiliation{\affiI}



\date{\today}

\begin{abstract}
We develop a self-consistent nonlinear extension of diffusive shock acceleration that incorporates cosmic-ray (CR) backreaction on the shock precursor together with a physically motivated upstream-escape mechanism that produces an exponential high-energy cutoff. The CR pressure gradient decelerates the upstream flow facing the shock wave, generating an extended precursor in which higher-rigidity particles sample a larger cumulative velocity gradient and thereby acquire a progressively harder spectrum. Finite-size/escape effects are modeled by a momentum-dependent loss term, which naturally terminates acceleration and steepens the spectrum near the cutoff. The precursor compression ratio is not imposed as a closure condition but is determined dynamically by requiring consistency between the injection rate inferred from thermal leakage at the subshock and the injection strength demanded by the nonlinear shock modification, with CR-driven wave heating providing stabilizing negative feedback. Applying the model to young supernova-remnant–like parameters and standard one-zone Galactic diffusion, we reproduce the main features of the latest DAMPE proton spectrum: gradual hardening from hundreds of GeV to multi-TeV energies and a subsequent exponential cutoff at tens of TeV. The resulting spectral evolution follows directly from the competition between precursor-mediated nonlinear feedback and upstream escape.

\end{abstract}

\maketitle

\section{Introduction}

Diffusive shock acceleration (DSA) is widely regarded as the primary mechanism for accelerating cosmic rays (CR) in astrophysical systems\citep{Blandford1987,Drury1983,Bell1978a,Bell1978b}. The steady-state particle distribution in such systems is typically described by the DSA equation. In the standard test-particle scenario, particles are assumed to gain energy only upon crossing the shock front, corresponding microscopically to the first-order Fermi acceleration process. Each shock crossing imparts a momentum gain proportional to the velocity difference across the shock, while there is also a finite probability that particles are advected downstream and lost from the acceleration region, resulting in the well-known power-law spectrum. A central assumption of the test-particle limit is that the accelerated particles exert no influence on the background plasma, that is, there is no feedback.

Building on earlier work \cite{1981APJ,McKenzie1982,1984APJ}, Malkov et al. \citep{Malkov1997,Malkov2000} demonstrated that non-thermal particles accelerated  by the shock can modify the shock structure, particularly in the upstream region. This feedback can induce a continuous velocity gradient in the upstream fluid due to the CR pressure gradient, forming a precursor. Within such a structure, particles can undergo acceleration throughout their upstream propagation, rather than solely at the shock front.
Because the diffusive-advective propagation distance of CRs scales as $D/u$, higher-energy particles can penetrate farther upstream before returning to the shock. Therefore, higher-energy particles experience a larger velocity gradient in the background flow, thus receiving greater acceleration, which leads to spectral hardening at the high-energy end—manifested as an energy-dependent deviation from a single power-law spectrum.

In Malkov’s formulation, closure of the nonlinear system is achieved by imposing a sharp upper momentum cutoff $p_1$, beyond which the CR distribution is abruptly truncated. While mathematically convenient, this assumption differs from the more physically motivated expectation of a gradual exponential cutoff (e-cut), which typically depends on the atomic number $Z$ and arises from finite acceleration timescales or escape through an upstream boundary. Moreover, key quantities such as the precursor compression ratio are selected via a mechanism of self-organized criticality that is intrinsically linked to the presence of this hard cutoff.

In this work, we improve upon Malkov’s approach in two key aspects. First, rather than imposing an ad hoc sharp cutoff at $p_1$, we model the high-energy termination of the CR spectrum using a physical exponential cutoff that naturally emerges from upstream escape, an unavoidable feature of realistic acceleration sites. Second, the precursor compression ratio is determined self-consistently from physically motivated boundary conditions, instead of being fixed by an externally imposed closure prescription. Within our framework, CR heating introduces a stabilizing negative feedback that regulates the nonlinear response of the system, contributes to downstream plasma heating, and ensures the existence of a well-behaved solution. By treating the upstream region as a constrained system bounded by the downstream thermal reservoir and the far-upstream medium, and by enforcing consistent injection conditions across the shock, the precursor compression ratio is implicitly fixed by the requirement that all physical boundary conditions be satisfied simultaneously.

Recent observations from DAMPE reveal a Z‑dependent e‑cut, accompanied by gradual spectral hardening from tens of GeV to several TeV before the cutoff. Several mechanisms could in principle produce spectra with both hardening and an exponential cutoff, including reacceleration by a local shock  \citep{Malkov2022,Malkov2021}, superposition of multiple source components \citep{Lv2025,PhysRevD.109.063001}, and self‑consistent halo models \citep{Chernyshov_2022}.

Motivated by these observations, we develop a feedback-driven acceleration model that self-consistently incorporates an e-cut through upstream escape and CR heating. Our numerical calculations show that this model can reproduce the latest DAMPE proton spectrum \citep{DAMPE2025} under conventional single-zone diffusion propagation in the Galaxy, thereby providing a physically motivated explanation for the observed spectral hardening followed by a cutoff.

This paper is organized as follows. Section~\ref{sec: method} outlines the methodology, Section~\ref{sec: results} presents a comparison between the model predictions and DAMPE observations and discusses the physical interpretation of the results, and Section~\ref{sec: summary} summarizes the main results.

\section{Method\label{sec: method}}

\subsection{Test Particle\label{sec: tp}}

Consider a plane shock propagating along the positive \(x\)-direction. In the shock rest frame, the upstream region corresponds to $x>0$, the downstream region to $x<0$, and the shock front is located at $x=0$. The phase-space distribution of CR particles is denoted by $f(x,p)$. The bulk plasma velocity in this frame is $u(x)$, directed opposite to the positive $x$-direction. In the following, we use dimensionless momentum normalized by the particle rest mass times \(c\), that is, \(p \rightarrow p/(mc)\), unless otherwise stated. The physical momentum is recovered by multiplying by \(mc\).

It is convenient to reformulate the problem in terms of the function $g(x,p)\equiv p^{3}f(x,p)$. Under this definition, $g(x,p)$ can be interpreted as the CR number density per logarithmic momentum interval, $\mathrm{d}N/\mathrm{d}\ln p = 4\pi g$. With these conventions, the standard DSA transport equation for $f(x,p)$ can be recast into the following form for $g(x,p)$~\citep{Malkov1997}:
\begin{equation}
\partial_x J(x,p) = \frac{p}{3}\,\frac{\mathrm{d}u}{\mathrm{d}x}\,\partial_p g(x,p),
\label{eq1}
\end{equation}
where
\begin{equation}
J(x,p) \equiv u(x)\,g(x,p) + D(p)\,\partial_x g(x,p).
\label{eq2}
\end{equation}
Here, $J(x,p)$ represents the combined convective and diffusive particle flux expressed in terms of $g(x,p)$, with the natural upstream boundary condition $J(\infty,p)=0$. Note that, by convention, this flux is directed opposite to the actual net particle flux in the upstream region. The right-hand side of Eq.~\eqref{eq1} describes particle acceleration due to compression of the background flow. Throughout this work, we assume a spatially homogeneous diffusion coefficient $D(p)$, depending only on momentum.

In the test-particle limit, the convection velocity gradient is confined to the shock front and can be written as
\(
\mathrm{d}u/\mathrm{d}x = \Delta u\,\delta(x),
\)
where \(\Delta u \equiv u_{0}-u_{2}\) denotes the velocity jump across the shock. Here, \(u_{0}\) is the convection velocity immediately upstream of the shock at \(x=0^{+}\), which, in the test-particle approximation, coincides with the constant flow velocity throughout the entire upstream region. The downstream flow velocity \(u_{2}\) is likewise taken to be constant throughout the downstream region.

Because the downstream region is sufficiently extended and contains an abundant population of thermal particles, the convective and diffusive fluxes there are aligned, preventing the buildup of a CR density gradient. As a result, the downstream solution reduces to the distribution at the shock front,
\(
g(x<0,p)=g_{0}(p)\equiv g(0^{+},p)=g(0^{-},p).
\)

We define the function 
\begin{equation}
\phi(x)=\int_{0}^{x}u(x')\mathrm{d}x',
\label{eq4}
\end{equation}
which represents the integrated flow velocity profile.

With these considerations, by combining Eq.~\eqref{eq1} and Eq.~\eqref{eq2}, the upstream distribution can be solved as
\begin{align}
g(x, p) & = e^{-\phi(x)/D(p)} \left[ g_{0}(p) + \int_{0^{+}}^{x} \frac{e^{\phi(x')/D(p)} \mathrm{d}x'}{D(p)} J(x', p) \right] \nonumber \\\
& = e^{-\phi(x)/D(p)} \left[ g_{0}(p) - \int_{0^{+}}^{x} \frac{e^{\phi(x')/D(p)} \mathrm{d}x'}{D(p)} \right. \nonumber \\\
&\qquad \left. \times \int_{x'}^{\infty} \frac{p\Delta u \delta({x}'')}{3}\partial_{p} g({x}'', p) ,\mathrm{d}{x}'' \right].
\label{eq3_intermediate}
\end{align}
At this stage, it is convenient to introduce the Green's function $G(x,x',p)$ defined by
\begin{equation}
G(x, x'; p) = -\frac{e^{-\phi(x)/D(p)}}{D(p)} \int_{0}^{\min(x, x')} e^{\phi(s)/D(p)}  \mathrm{d}s.
\label{eq_green}
\end{equation}
The solution  can be rewritten in the compact form
\begin{align}
g(x,p) = &g_{0}(p) e^{-\phi(x)/D(p)} \nonumber\\\
&+\int_{0^{+}}^{\infty} G(x,\xi,p) \frac{p\Delta u \delta(\xi)}{3}  \partial_{p} g(\xi,p) \mathrm{d}\xi.
\label{g_distribution}
\end{align}

In the test-particle approximation, the integral term in Eq.~\eqref{g_distribution} vanishes. Consequently, the upstream distribution reduces to
\begin{equation}
g(x, p) = g_{0}(p) e^{-\phi(x)/D(p)}.
\label{eq3}
\end{equation}

To determine the momentum dependence at the shock, the equation governing $g_{0}(p)$ is obtained by integrating Eq.~\eqref{eq1} across a narrow interval containing the shock front:
\begin{equation}
\Delta u \, g_{0} + D \, \partial_{x} g\big|_{x=0^{+}} = \frac{p \Delta u}{3} \partial_{p} g_{0}
\label{eq5}
\end{equation}

Substituting the upstream solution Eq.\eqref{eq3} for $g(x,p)$ into Eq.~\eqref{eq5} yields
\begin{equation}
-u_{2} g_{0} = \frac{p \Delta u}{3} \partial_{p} g_{0}
\label{eq6}
\end{equation}

This equation expresses the balance between the flux of particles advected downstream to infinity and the momentum gain associated with compression at the shock front. Its solution is a power-law spectrum,
\begin{equation}
g_{0}(p) = g_{0}(p_{\text{inj}}) \left( \frac{p}{p_{\text{inj}}} \right)^{-\frac{3}{r_{s} - 1}}
\label{eq7}
\end{equation}
where $r_{s}\equiv u_{0}/u_{2}$ is the shock compression ratio and $p_{\mathrm{inj}}$ denotes the injection momentum. The corresponding spectral index for the phase-space distribution $f_{0}(p)$ is the well-known result $-3r_{s}/(r_{s}-1)$.


In the test-particle regime of diffusive shock acceleration, the upstream region is effectively infinite, such that convection always returns particles to the shock regardless of their energy, resulting in a spectrum without a high-energy cutoff. In realistic acceleration environments, the upstream region has a finite spatial extent, and particle transport is governed by the competition between diffusion and convection. Once a particle’s diffusion length exceeds the characteristic size of the system, it is likely to escape upstream before being advected back to the shock. As we will show below, this escape process naturally leads to an exponential cutoff (e-cut) in the energy spectrum. Two complementary approaches are commonly employed to model such finite-system effects:
\begin{enumerate}
\item Absorbing-boundary model: A physical absorbing boundary is imposed at a fixed distance $L$ upstream. Particles reaching $x=L$ are instantaneously removed from the system, thereby defining an explicit maximum diffusion scale~\citep{1999APJ,Berezhko_1988}.
\item Escape-term model: A momentum-dependent escape term, $-f/\tau_{\mathrm{esc}}$, is introduced into the transport equation. The escape rate $\tau_{\mathrm{esc}}^{-1}=D/\xi^{2}$ encapsulates the notion that particles with larger diffusion coefficients (correspond to higher momenta) escape more efficiently from a region of characteristic size $\xi$. This formulation directly captures the competition between diffusion and escape processes~\citep{Schlickeiser2002,Drury1999,Protheroe1999,Fukugita1994}.
\end{enumerate}

In this work, we adopt the latter approach. Accordingly, Eq.~\eqref{eq1} is modified to
\begin{equation}
\partial_{x} J - \tau_{\text{esc}}^{-1} g = \frac{p}{3} \frac{\mathrm{d}u}{\mathrm{d}x} \partial_{p} g
\label{eq8}
\end{equation}

while the corresponding solution for $g(x,p)$, replacing Eq.~\eqref{g_distribution}, becomes
\begin{equation}
g = e^{-\phi/D} \left[ g_{0} + \int_{0^{+}}^{\infty} G(x,\xi,p)\tau_{\text{esc}}^{-1} g(\xi,p) \, d\xi \right]
\label{eq9}
\end{equation}

In the test-particle limit, the solution of Eq.~\eqref{eq9} takes the form
\( g(x,p)=g_{0}(p)\,e^{-k(p)x} \) with \(k>0\).
Substituting this ansatz into Eq.~\eqref{eq9} yields the condition that $k$ must satisfy
\begin{equation}
D k^{2} - u_{0} k - \tau_{\text{esc}}^{-1} = 0
\label{eq10}
\end{equation}
which admits the solution
\begin{equation}
k = \frac{u_{0} \left( 1 + \sqrt{1 + 4 \left( \frac{v_{\text{esc}}}{u_{0}} \right)^{2}} \right)}{2 D}
\label{eq11}
\end{equation}
Here, \(v_{\mathrm{esc}}(p)\equiv D/\xi\) can be interpreted as an effective escape velocity associated with diffusive transport.

Substituting this result into Eq.~\eqref{eq5}, one obtains the particle spectrum at the shock,
\begin{multline}
g_{0}(p) = g_{0}(p_{\text{inj}}) \left( \frac{p}{p_{\text{inj}}} \right)^{-\frac{3}{r_{s} - 1}} \\
\times \exp\left( \frac{-3 r_{s}}{r_{s} - 1} \int_{p_{\text{inj}}}^{p} 
\left[ \frac{\sqrt{1 + 4 \left( \frac{v_{\text{esc}}(p')}{u_{0}} \right)^{2}} - 1}{2 p'} \right] dp' \right)
\label{eq12}
\end{multline}
It can be seen that when $v_{\mathrm{esc}}\sim u_{0}$, the escape effect becomes significant and the spectrum begins to deviate from a pure power law. In the high-energy limit \(v_{\mathrm{esc}}\gg u_{0}\), the integrand in the exponential grows approximately as \(v_{\mathrm{esc}}/u_{0}\), giving rise to a pronounced exponential cutoff in the spectrum.

\subsection{Feedback\label{sec: feedback}}
In the presence of CR acceleration, the upstream diffusion--convection balance produces a CR pressure gradient that, in turn, drives a spatial gradient in the background flow, resulting in the upstream velocity $u(x)$ position dependent. Starting from Eq.~\eqref{eq8} and following the procedure of Eq.~\eqref{g_distribution}, the upstream distribution $g(x,p)$ satisfies:
\begin{equation}
g(x, p) = g^{(0)}(x, p) + W[g](x, p),
\label{eq16}
\end{equation}

The functional $W[g]$ accounts for the effects of compression and particle escape and is given by
\begin{equation}
W[g](x, p) = \int_{0^{+}}^{\infty} G(x, x'; p) S[g](x', p) \, \mathrm{d}x'.
\label{eq17}
\end{equation}
The source term $S[g](x,p)$ includes the acceleration effect and particle escape effect,
\begin{equation}
S[g](x, p) = \frac{p}{3} \frac{\mathrm{d}u}{\mathrm{d}x} \partial_{p} g + \tau_{\text{esc}}^{-1} g.
\label{eq18}
\end{equation}
Here the acceleration term depends on the velocity‑gradient distribution that exists throughout the upstream precursor and is self‑consistently driven by cosmic rays, so acceleration occurs not only at the shock front but also across the entire upstream region.

The zeroth-order solution $g^{(0)}(x,p)$ corresponds to pure advection--diffusion in the absence of sources and is given by
\begin{equation}
g^{(0)}(x, p) = g_{0}(p) e^{-\phi(x)/D(p)}.
\label{eq20}
\end{equation}
According to Eq.~\eqref{eq16}, the upstream distribution can be expressed as a series expansion around the zeroth-order solution $g^{(0)}$. In the main value region of $g^{(0)}$ which satisfies $xu/D \ll 1$ and $v_{\mathrm{esc}}/u \ll 1$, its contribution dominates the feedback on the flow field. In this regime, the first-order approximation can be readily evaluated as

\begin{equation}
g^{(1)} = g^{(0)} e^{W[g^{(0)}]/g^{(0)}} \simeq g_{0} e^{ -\frac{q \phi}{3 D} - \tau_{\text{esc}}^{-1} \int_{0}^{x} \frac{\mathrm{d}x'}{u(x')} }
\label{eq21}
\end{equation}
where $q(p)\equiv -\partial\ln f_{0}/\partial\ln p
= 3-\partial\ln g_{0}/\partial\ln p$ denotes the spectral index. In deriving Eq.~\eqref{eq21}, higher-order terms proportional to $D^{-1}$ have been neglected.

Noting that $q/3>1 \gg (v_{\mathrm{esc}}/u)^{2}$, the upstream distribution may be approximated as
\begin{equation}
g(x, p) \approx g_{0}(p) e^{ -\frac{q(p) \phi(x)}{3 D(p)} }
\label{eq22}
\end{equation}

For computational simplicity, higher-order corrections associated with the truncation have been omitted. As discussed in Appendix~A, such corrections tend to suppress the feedback strength of the pre-truncation system. Nevertheless, numerical studies by Caprioli~\citep{Caprioli2010} indicate that incorporating higher-order truncation effects into the feedback does not substantially modify the spectral shape below the cutoff. Instead, the dominant impact of truncation manifests primarily in the high-energy region beyond the cutoff.

With the inclusion of escape and feedback effects, Eq.~\eqref{eq5} is modified to
\begin{equation}
u_{2} g_{0} + \frac{p \Delta u}{3} \partial_{p} g_{0} + \int_{0^{+}}^{\infty} S[g] \, \mathrm{d}x = 0
\label{eq23}
\end{equation}

Using the first-order approximation $g^{(1)}$, the combined effects of escape and nonlinear feedback can be cast into a closed equation for the momentum-dependent spectral index $q(p)$,
\begin{equation}
q(p) = 3 + \frac{3}{R r_{s} U(p)} + \frac{1}{q(p) U(p) \alpha} \left( \frac{3v_{\text{esc}}}{u_{1}} \right)^{2} + \frac{\mathrm{d} \ln U}{\mathrm{d} \ln p}.
\label{eq24}
\end{equation}
Where the subscript $1$ denotes quantities evaluated at the far-upstream boundary. $R\equiv u_{1}/u_{0}$ denotes the compression ratio across the entire upstream precursor, while $Rr_{s}=u_{1}/u_{2}$ is the total compression ratio of the system. The parameter $\alpha\equiv\bar{u}/u_{1}\in[R^{-1},1]$ characterizes the effective upstream flow speed sampled by the particles. To quantify the escape effect, we introduce a cutoff parameter $\chi$ defined at the reference momentum $p=1$ (mc) as 

\begin{equation}
\chi \equiv \frac{\sqrt{\alpha}\, u_1}{3\,v_{\mathrm{esc}}(1)}.
\label{eq:chi}
\end{equation}
A larger value of \(\chi\) delays the onset of the exponential cutoff. Assuming the diffusion coefficient of the power-law follows $D(p)\propto p^{\beta}$ we can rewrite the escape term and q(p) using $\chi$:
\begin{equation}
q(p) = 3 + \frac{3}{R r_{s} U(p)} + \frac{1}{q(p) U(p)} \left( \frac{p^{\beta}}{\chi} \right)^{2} + \frac{\mathrm{d} \ln U}{\mathrm{d} \ln p}.
\label{eq24-1}
\end{equation}

If we calculate $q(p)$, we can use it to calculate $g_{0}(p)$:

\begin{equation}
    g_{0}(p)=g_{0}(p_{inj})e^{-\int_{p_{inj}}^{p}\frac{q\left({p}'\right)-3}{{p}'}d{p}'}
    \label{eq29}
\end{equation}

The cumulative normalized velocity gradient experienced by particles of momentum $p$ is defined as
\begin{equation}
U(p) \equiv \frac{\Delta u + V(p)}{u_{1}} = \frac{r_{s} - 1}{R r_{s}} + \frac{1}{u_{1}} V(p),
\label{eq25}
\end{equation}
where $V(p)$ represents the weighted contribution from the spatially distributed velocity gradient,
\begin{equation}
\begin{aligned}
V(p)
=\, & \frac{1}{g_{0}}
\int_{0^{+}}^{\infty}
\frac{\mathrm{d}u}{\mathrm{d}x}\,g\,\mathrm{d}x \\
=\, &
\int_{0^{+}}^{\infty}
\frac{\mathrm{d}u}{\mathrm{d}x}
\exp\!\left[
-\frac{q(p)\,\phi(x)}{3D(p)}
\right]
\,\mathrm{d}x .
\end{aligned}
\label{eq26}
\end{equation}

In order to obtain the spectral index $q(p)$, we need to know the information of the flow velocity gradient $\frac {du}{dx}$ to calculate $V(p)$. Then, the flow field distribution is self consistently driven by CR pressure, following the standard mass and momentum conservation equations \citep{Axford1977}:
\begin{equation}
\rho(x) u(x) = \rho_{1} u_{1},
\label{eq13}
\end{equation}
\begin{equation}
\rho(x) u^{2}(x) + P_{\text{CR}}(x) + P_{g}(x) = \rho_{1} u^{2}_{1}  + P_{g,1},
\label{eq14}
\end{equation}
where $P_g$ represents thermal pressure. The CR pressure is given by

\begin{equation}
    P_{\mathrm{CR}}(x) = \frac{mc}{3}\int_{p_{\mathrm{inj}}}^{\infty} p v \, \mathrm{d}N = \frac{4\pi m c^{2}}{3}\int_{0}^{\infty}\frac{p\,g(x,p)}{\sqrt{p^{2}+1}}\,\mathrm{d}p, 
    \label{PCR}
\end{equation}
$v$ is the velocity of particles in the laboratory system and we have used relativistic kinematics for convenience. It is useful to separate the overall normalization from the momentum-dependent shape by writing
\begin{equation}
g(x,p) = g_{0}(p_{\mathrm{inj}})\,\tilde{g}(x,p),
\label{eq:tildeg}
\end{equation}
with \(\tilde{g}(x,p)\) normalized such that \(\tilde{g}(0,p_{\mathrm{inj}})=1\). The quantity \(g_{0}(p_{\mathrm{inj}})\) is the shock-frame distribution at the injection momentum; it has dimensions of number density. The combination \(\frac{4\pi}{3} m c^2 g_0(p_{\mathrm{inj}})\) then has dimensions of pressure and provides a characteristic scale for the CR pressure contributed by the injected particles. Physically, it corresponds to the CR pressure that would result if all injected particles were concentrated at the injection momentum with a phase-space density \(f_0(p_{\mathrm{inj}})\) and treated in the relativistic limit (where \(p \gg mc\)). In reality, the full CR pressure is obtained by integrating over the entire momentum distribution, but it scales linearly with this characteristic pressure, with the proportionality factor determined by the shape of the spectrum. Based on this, we introduce a dimensionless injection rate parameter:
\begin{equation}
    \eta \equiv \frac{4\pi}{3}\frac{m c^{2}}{\rho_1 u_1^{2}}\,g_0(p_{\mathrm{inj}}), 
    \label{eta_def}
\end{equation}
which quantifies the ratio of this characteristic CR pressure to the incoming ram pressure \(\rho_1 u_1^2\). Physically, \(\eta\) directly controls the strength of the nonlinear feedback: the actual CR pressure scales as \(P_{\mathrm{CR}} \propto \eta \times (\rho_1 u_1^2)\). More explicitly, using the definition of \(\eta\) and the normalized distribution \(\tilde{g}\), the CR pressure at any position \(x\) can be written as
\begin{equation}
P_{\mathrm{CR}}(x) = \eta \rho_1 u_1^2 \int_0^\infty \frac{p\tilde{g}(x,p)}{\sqrt{p^2+1}}  \, dp,
\label{eq:PCR_eta}
\end{equation}
which shows that \(\eta\) sets the overall normalization of the CR pressure relative to the ram pressure. If no particles are injected, \(\eta = 0\) and consequently \(g_{0}(p_{\mathrm{inj}})=0\), then Eq.~\eqref{eq:PCR_eta} gives \(P_{\mathrm{CR}}(x)=0\) everywhere. In that case the momentum conservation Eq.~\eqref{eq14} reduces to \(\rho u^{2}+P_{g}= \rho_{1}u_{1}^{2}+P_{g,1}\). Together with the adiabatic relation \(P_{g}\propto \rho^{\gamma}\) and mass conservation Eq.~\eqref{eq13}, this algebraic equation forces \(u(x)=u_{1}\) for all \(x\): the upstream flow remains unperturbed and no feedback operates.

In the presence of injection (\(\eta>0\)), the CR pressure modifies the flow. To proceed, we neglect the thermal pressure \(P_{g}\) in the momentum balance when evaluating the velocity gradient; this is justified in the strong-shock regime where the ram pressure dominates and the CR pressure provides the leading correction. Using Eqs. \eqref{eq13} and \eqref{eq14} together with \eqref{eq:tildeg}, one obtains an expression for \(\mathrm{d}u/\mathrm{d}x\) that depends on the CR distribution. Substituting this into the definition Eq.~\eqref{eq26} of \(V(p)\) and employing the first-order approximation Eq.~\eqref{eq22} for \(g(x,p)\) yields

\begin{equation}
\begin{aligned}
V(p)
=\, & u_{1}\,\eta
\int_{p_{\mathrm{inj}}}^{\infty}
\frac{ p'\,\frac{q(p')}{3D(p')}\,
\tilde{g}_{0}(p') }
{\sqrt{p'^{2}+1}}
\,\mathrm{d}p' \\
& \times
\int_{0}^{\infty}
\exp\!\left[
-\Bigl(
\frac{q(p')}{3D(p')}
+\frac{q(p)}{3D(p)}
\Bigr)\phi
\right]
\,\mathrm{d}\phi ,
\end{aligned}
\label{eq27}
\end{equation}
which can be evaluated analytically to give
\begin{equation}
V(p)
= u_{1}\,\eta
\int_{p_{\mathrm{inj}}}^{\infty}
\frac{ p'\,\tilde{g}_{0}(p')\,\mathrm{d}p' }
{\sqrt{p'^{2}+1}}
\frac{ q(p')\,D(p) }
{ q(p)\,D(p') + q(p')\,D(p) }.
\label{eq28}
\end{equation}

Equation \eqref{eq28} shows explicitly that the cumulative velocity-gradient measure \(V(p)\) is proportional to the injection rate \(\eta\). Hence \(\eta\) acts as the strength parameter of the nonlinear feedback: a larger \(\eta\) means more particles are injected, leading to stronger CR pressure and a more pronounced modification of the upstream flow. In the limit \(\eta\to0\), \(V(p)\to0\) and the system reverts to the test-particle description with no spectral hardening. This linear scaling of \(V\) with \(\eta\) also implies that the entire nonlinear response of the shock structure is controlled by the single dimensionless parameter \(\eta\), which itself is determined by the microphysics of injection as described in Sec.~\ref{sec: injection}.

\subsection{Injection\label{sec: injection}}

The acceleration framework introduced in Section~\ref{sec: feedback} is specified by a set of key parameters: the precursor compression ratio \(R\), the subshock compression ratio \(r_s\), the injection rate \(\eta\), the injection momentum \(p_{\text{inj}}\), the diffusion coefficient \(D(p)\), and the far-upstream Mach number \(M \equiv u_1 / c_{s,1}\). Most of these quantities are not free parameters but are instead constrained by boundary conditions and fundamental physical relations.

In general, particles injected into the acceleration process originate from the high-energy tail of the downstream thermal distribution, where their momenta exceed the thermal momentum by several factors \citep{Kang2002,Berezhko1995}. Only such suprathermal particles possess a non-negligible probability of recrossing the shock and entering the diffusive acceleration cycle. The injection momentum is therefore parametrized as
\(p_{\text{inj}} = \zeta p_{\text{th},2}\), where the dimensionless parameter \(\zeta\) (typically in the range \(2\!-\!4\)) specifies the threshold, in units of the downstream thermal momentum \(p_{\text{th},2}\), required for injection. This yields
\begin{equation}
p_{\text{inj}} = \zeta p_{\text{th},2}= \zeta p_{\text{th},1} \sqrt{ \frac{T_2}{T_0} \frac{T_0}{T_1} } ,
\label{eq30}
\end{equation}
where the far-upstream thermal momentum is defined as \(p_{\text{th},1} = \sqrt{2 m k_B T_1}\). The temperature jump \(T_2/T_0\) across the subshock follows from the Rankine--Hugoniot relations, while upstream compression gives
\begin{equation}
\frac{T_0}{T_1} = \frac{P_{g,0} / \rho_0}{P_{g,1} / \rho_1} .
\label{eq31}
\end{equation}

Under purely adiabatic compression, the thermal gas pressure obeys \(P_g \propto \rho^{\gamma}\), with \(\gamma\) the adiabatic index of the background plasma. In the non-adiabatic case, however, additional heating arises from CR-induced wave excitation in the precursor. The contribution of the wave energy to the gas heating can be described by the first law of thermodynamics \citep{McKenzie1982,Berezhko1995},
\begin{equation}
\frac{d}{dx} \left[ \frac{1}{2} \rho u^3 + P_g u \frac{\gamma}{\gamma - 1} \right]
+ \left( u - v_s \right) \frac{d}{dx} P_{\text{CR}} = 0 ,
\label{eq32}
\end{equation}
which, upon combining Eq.~\eqref{eq13} and Eq.~\eqref{eq14}, can be recast as
\begin{equation}
\begin{aligned}
\frac{d}{dx} \left( P_g \rho^{-\gamma} \right)
&= \frac{(\gamma - 1) v_s}{u} \rho^{-\gamma} \partial_x P_{\text{CR}} \\
&\approx - \rho_1 u_1 \frac{(\gamma - 1) v_s}{u} \rho^{-\gamma} \partial_x u .
\end{aligned}
\label{eq33}
\end{equation}
Here, \(v_s\) denotes the speed of backward-propagating waves excited in the upstream medium, with \(v_s \propto u^{s}\); the choice \(s=1/2\) typically corresponds to Alfvén-wave heating(This is our default situation).

The resulting gas pressure profile in the precursor can then be written as
\begin{equation}
P_g(x) = P_{g,1} \left( \frac{u_1}{u} \right)^{\gamma} \left[ 1 + H(x) \right] ,
\label{eq34}
\end{equation}
where the non-adiabatic correction factor is
\begin{equation}
H(x) = \frac{\gamma (\gamma - 1)}{\gamma + s} \frac{M^2}{M_s}
\left[ 1 - \left( \frac{u}{u_1} \right)^{\gamma + s} \right] ,
\label{eq35}
\end{equation}
and \(M_s \equiv u_1 / v_{s,1}\). In the limit \(M_s \to \infty\), the correction term vanishes (\(H \to 0\)), and the adiabatic result is recovered.

With these ingredients, the injection momentum can be expressed as
\begin{equation}
p_{\text{inj}} = \zeta p_{\text{th},1}
\sqrt{
R^{\gamma - 1} (1 + H_0)
\frac{\gamma + 1 - (\gamma - 1) r_s^{-1}}
{\gamma + 1 - (\gamma - 1) r_s}
} ,
\label{eq36}
\end{equation}
where \(H_0 \equiv H(x=0)\). The subshock compression ratio \(r_s\) follows from momentum conservation and the Rankine--Hugoniot conditions, yielding
\begin{equation}
r_s = \frac{\gamma + 1}
{\gamma - 1 + 2 (1 + H_0) R^{\gamma + 1} M^{-2}} .
\label{eq37}
\end{equation}

As discussed in Section~\ref{sec: feedback}, the injection rate \(\eta\) controls the strength of nonlinear feedback in the system. From a physical perspective, \(\eta\) is related to the leakage of particles from the downstream thermal pool into the non-thermal population. It can be written Eq.~\eqref{eta_deifi} as
\begin{equation}
\eta = \frac{4\pi}{3} \frac{m c^2}{\rho_1 u_1^2} g_0(p_{\text{inj}})
= \frac{4 R r_s}{3 \sqrt{\pi}} \frac{c^2}{u_1^2} \zeta^3 e^{-\zeta^2}.
\label{eq40}
\end{equation}
In the last step, we have imposed continuity between the thermal and non-thermal distributions at \(p_{\text{inj}}\), that is, \(f_0(p_{\text{inj}}) = f_{\text{th},2}(p_{\text{inj}})\), and substituted the downstream thermal spectrum. For given \(\zeta\), \(M\), and \(p_{\text{th},1}\) (or equivalently \(T_1\)), this relation uniquely determines \(\eta\).

From a complementary viewpoint, \(\eta\) also quantifies the injection strength required to sustain a precursor compression ratio \(R\) in the nonlinear regime. Using Eq.~\eqref{eq13}--\eqref{eq14} together with the boundary condition \(R = u_0 / u_1\), one obtains
\begin{equation}
\eta =
\left(1 - R^{-1} - \dfrac{R^{\gamma}(1+H_{0}) - 1}{\gamma M^{2}}\right)\left(\int_{0}^{\infty}\frac{p\tilde{g}_{0}(p)dp}{\sqrt{p^2+1}}\right)^{-1}.
\label{eq41}
\end{equation}

The physically realized precursor compression ratio \(R\) is determined by the requirement that these two independent estimates  of the injection rate coincide(Eq.~\eqref{eq41} and ~\eqref{eq40}), ensuring consistency between the microphysical injection process and the global nonlinear modification of the shock.

\section{Results\label{sec: results}}

To enable a direct comparison with DAMPE observations, the source spectrum must be consistently connected to the propagated spectrum at Earth. Within the conventional one-zone diffusion–propagation framework, the observed differential flux in kinetic energy can be written as

\begin{equation}
\frac{dN}{dE_{\text{k}}} \propto (E_{\text{k}} + m_p c^2)
\left( \frac{p(E_{\text{k}})}{m_p c} \right)^{-2-\delta}
\tilde{g}_0 \left( \frac{p(E_{\text{k}})}{m_p c} \right) ,
\label{eq42}
\end{equation}

where the prefactor $(E_{\text{k}} + m_p c^2)$ arises from the transformation between momentum and kinetic-energy space, and $m_p$ denotes the proton mass. In addition, here we revert to using physical momentum with dimensions. The second factor accounts for CR propagation in the interstellar medium, which is approximated as a softening of the source power-law index by $p^{-\delta}$. Here the diffusion coefficient is parameterized as $D_{\text{ISM}} \propto p^{\delta}$, with $\delta = 0.5$ fixed by fits to the boron-to-carbon (B/C) ratio, consistent with previous studies \citep{Yuan2019,Derome2019,PhysRevD.99.123028,PhysRevD.95.083007}. The final factor, $\tilde{g}_0$, represents the injected source spectrum computed from Eq.~\eqref{eq29}. This assumes that the observed CRs correspond to the particles accumulated downstream during the active acceleration phase and released when the shock wave dissipates.

\begin{figure}[H]
\centering
\includegraphics[width=0.45\textwidth]{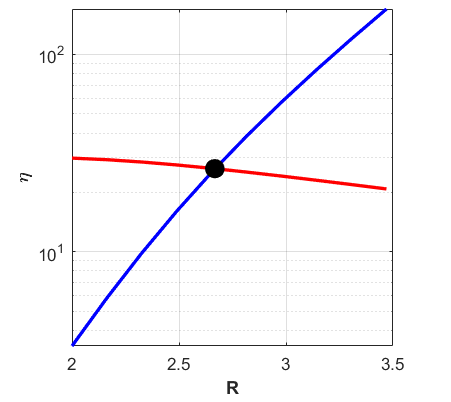} 
\caption{Injection-rate response curves: the thermal-injection constraint (red, Eq.~\eqref{eq40}) and the system's self-consistent nonlinear response $\eta(R)$ (blue, Eq.~\eqref{eq41}). Their intersection determines the precursor compression ratio $R$. }
\label{fig:2}
\end{figure}

The self-consistent determination of the precursor compression ratio $R$ is illustrated in Fig.~\ref{fig:2}. Within the precursor, a larger $R$ corresponds to stronger nonlinear feedback from CRs, implying enhanced shock modification and, consequently, a higher injection rate required to sustain the system. This behavior is captured by the blue curve in Fig.~\ref{fig:2}, which represents the nonlinear response $\eta(R)$. Independently, the injection rate inferred from thermal leakage (red curve) provides a second constraint. The physically admissible solution is given by the intersection of these two curves, yielding a unique $R$ for a given set of parameters.

An important feature of the nonlinear response curve is the presence of a distorted (oscillatory) region for intermediate values of $R$. This distortion arises from the delicate balance between the positive feedback of particle injection and the negative feedback of plasma heating induced by CR-excited waves. The positive feedback—controlled by the injection parameter $\zeta$—tends to increase $R$ by enhancing the CR pressure gradient, while the negative feedback—governed by the wave-heating parameter $M_s$—suppresses $R$ by raising the upstream gas pressure and reducing compressibility. When these two effects are not properly matched, the $\eta(R)$ curve can develop unphysical oscillations or even fail to intersect the thermal-injection curve, leading to either multiple solutions or no solution at all. In practice, avoiding such pathological behavior requires a balanced pairing: a relatively high injection (small $\zeta$) must be accompanied by strong heating (small $M_s$), and vice versa. This matching condition ensures that the system remains in a stable, physically realizable regime where the two curves intersect at a reasonable $R$, typically confined to the range $2 \lesssim R \lesssim 5$.

\begin{figure}[H]
\centering
\includegraphics[width=0.45\textwidth]{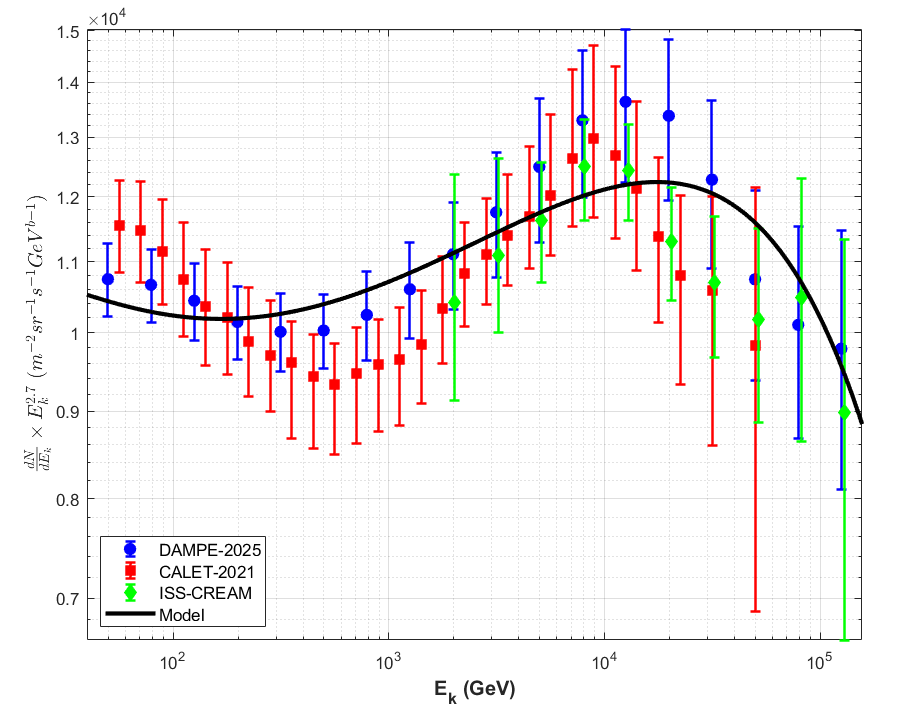} 
\caption{Comparison between the theoretical model (black curve) and the latest DAMPE-2025 \citep{DAMPE2025} proton spectrum, together with CALET-2022 \citep{Adriani2022} and ISS-CREAM \citep{Choi2022} data. }
\label{fig:1}
\end{figure}
With the above insights, we now apply the model to interpret the latest DAMPE proton spectrum. Adopting parameters characteristic of an early-phase supernova remnant environment with a strong shock ($M = 200$) and a relatively thermal upstream medium ($p_{\mathrm{th},1} = 0.0005\,m_p c$, corresponding to $T_1 \sim 10^6\,\mathrm{K}$), we fix the diffusion coefficient in the source region to follow $D(p) \propto p^{1/3}$, which is the Kolmogorov scaling \citep{Kolmogorov1941}. The injection parameter $\zeta = 2.3$, the wave-heating parameter $M_s = 14$, and the cutoff parameter $\chi \approx 43$ (defined in Eq.~\eqref{eq:chi}) are chosen such that the thermal-injection and nonlinear-response curves intersect cleanly, yielding a precursor compression ratio $R \approx 2.7$ and a subshock compression ratio $r_s \approx 1.685$. The resulting source spectrum, after propagation through the Galaxy with the standard diffusion index $\delta = 0.5$, is shown in Fig.~\ref{fig:1} (black curve) alongside the latest DAMPE data and other measurements. The model successfully reproduces the gradual hardening from several hundred GeV to multi-TeV energies and the subsequent exponential cutoff above $\sim 20$ TeV, with a change in spectral index $\Delta q \sim 0.2$.

The slight mismatch at the lowest energies—the predicted hardening onset appears about $100$GeV lower than indicated by the data—may be attributed to the neglect of thermal pressure in the calculation of the velocity gradient $V(p)$ (in Sec.~\ref{sec: feedback}). This simplification tends to overestimate $\mathrm{d}u/\mathrm{d}x$ in the precursor, thereby shifting the hardening to slightly lower energies, but does not alter the essential physics.

\begin{figure}[H]
\centering
\includegraphics[width=0.45\textwidth]{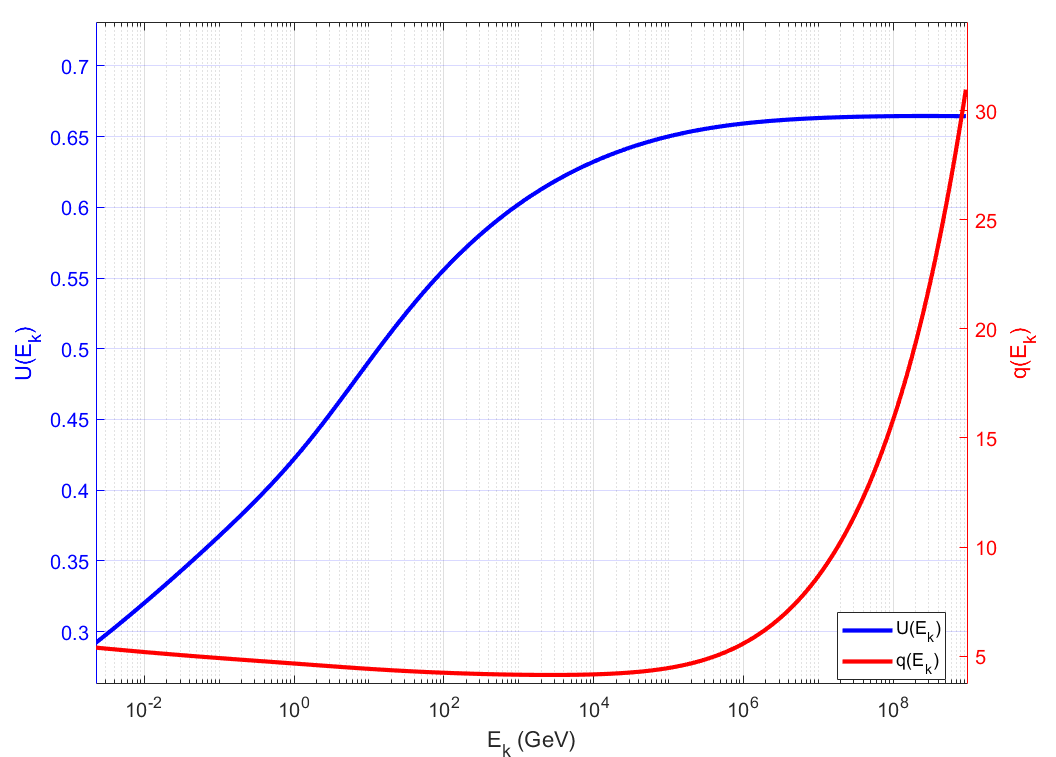} 
\caption{Cumulative normalized velocity gradient $U$ (blue) and spectral index $q$ (red) as functions of kinetic energy, computed for the same parameters as in Fig.~\ref{fig:1}.}
\label{fig:3}
\end{figure}

The spectral hardening and subsequent cutoff observed in Fig.~\ref{fig:1} can be understood by examining the energy-dependent behavior of the cumulative normalized velocity gradient $U$ and the spectral index $q$, shown in Fig.~\ref{fig:3}. In the feedback-dominated region below $\sim 10^{4}$~GeV, a clear anti-correlation between $U$ and $q$ is evident. This behavior arises because higher-energy particles, with their larger diffusion coefficients, propagate farther upstream and experience a greater integrated velocity gradient $V(p)$, as described by Eqs.~\eqref{eq25} and \eqref{eq28}. Consequently, $U$ increases with energy, and through the relation encoded in Eq.~\eqref{eq24-1}, the spectral index $q$ decreases accordingly, leading to the gradual hardening seen in the propagated spectrum.

A transition occurs near $\sim 10^{4}$~GeV, beyond which escape effects become dominant. At these energies, particles escape from the upstream region before returning to the shock, halting further accumulation of the velocity gradient. Consequently, $U$ saturates to an asymptotic value, while the spectral index $q$ increases sharply due to the escape term in Eq.~\eqref{eq24-1}, directly giving rise to the exponential cutoff in the spectrum. This competition between feedback-driven hardening and escape-induced softening naturally accounts for the observed spectral shape.

\section{Summary\label{sec: summary}}

In this work, we develop and investigate a self-consistent, nonlinear model of diffusive shock acceleration that explicitly incorporates CR feedback and particle escape. Going beyond the standard test-particle framework, we extend the approach of Malkov \emph{et al.} by introducing two key improvements. First, we implement a physically motivated exponential cutoff (e-cut) in the momentum spectrum, arising naturally from finite upstream boundaries or escape, rather than adopting an \emph{ad hoc} sharp truncation. Second, we determine the precursor compression ratio in a physically self-consistent manner by matching the injection rate inferred from thermal leakage with the nonlinear response required by the system. Within this framework, cosmic-ray induced wave heating is identified as an essential negative feedback mechanism that counterbalances the positive feedback from particle injection, stabilizes the shock structure, and guarantees the existence of physically admissible solutions.

We apply this framework to interpret the latest DAMPE measurements of the CR proton spectrum. Our numerical results show that, within a conventional single-zone diffusion model and for parameters characteristic of young SNRs, the model successfully reproduces the observed gradual spectral hardening from hundreds of GeV to several TeV, followed by an exponential cutoff at higher energies. The resulting spectral shape emerges naturally from the energy-dependent coupling between particle transport and the shock structure modified by CR pressure in the precursor.

We note that, for numerical tractability, certain higher-order effects were neglected in the present formulation, particularly refinements related to particle escape and the detailed contribution of thermal pressure in the vicinity of the cutoff. These simplifications may account for minor discrepancies between the model predictions and the data, such as the slightly earlier onset of spectral hardening. A more complete treatment of these effects is deferred to future work.

\acknowledgments
This work is supported by the National Key R\&D program of China under the grant 2024YFA1611402, the National Natural Science Foundation of China under Grants No. 12175248, No. 12105292, and No. 12393853. 

\bibliography{apssamp}
\appendix
\section*{Appendix A}

We now examine in detail the effect of escape on nonlinear feedback acceleration. We know that for the operator

\begin{equation}
L_1[g] = \partial_x (u g + D \partial_x g)
\label{eq43}
\end{equation}

there is a Green's function

\begin{equation}
G_1(x, \xi) = -e^{-\phi(x)/D} \int_{0}^{\min(x, \xi)} e^{\phi(s)/D} \, \frac{ds}{D}
\label{eq44}
\end{equation}

satisfying the boundary conditions \(G(0, \xi) = G(\infty, \xi) = 0\), where we omit the explicit \(p\) dependence.

Now for the operator \(L_2 = L_1 - \tau_{\text{esc}}^{-1}\) and its Green's function \(G_2(x, \xi)\) satisfying:

\begin{equation}
L_{2,x} G_2(x, \xi) = \delta(x - \xi)
\label{eq45}
\end{equation}

\begin{equation}
G_2(0, \xi) = G_2(\infty, \xi) = 0
\label{eq46}
\end{equation}

where \(L_{2,x}\) denotes acting on \(x\), we have:

\begin{equation}
L_{1,x} \left( G_2(x, \xi) - G_1(x, \xi) \right) = \tau_{\text{esc}}^{-1} G_2(x, \xi)
\label{eq47}
\end{equation}

This is equivalent to the following integral equation:

\begin{equation}
G_2(x, \xi) = G_1(x, \xi) + \tau_{\text{esc}}^{-1} \int_0^{\infty} G_1(x, \eta) G_2(\eta, \xi) \, d\eta
\label{eq48}
\end{equation}

Before the cutoff, where \(\frac{v_{\text{esc}}}{u} \ll 1\), treating escape as a perturbation, we estimate the first-order behavior:

\begin{equation}
G_2^{(1)}(x, \xi) = G_1(x, \xi) + \tau_{\text{esc}}^{-1} \int_0^{\infty} G_1(x, \eta) G_1(\eta, \xi) \, d\eta
\label{eq49}
\end{equation}

Now, up to order \(D^{-1}\), we estimate:

\begin{equation}
\small
\begin{aligned}
g^{(1)}(x,p) &\simeq g^{(0)}(x,p) + \int_{0^+}^{\infty} \frac{3}{p} \frac{du}{d\xi} G_2^{(1)}(x,\xi';p) \partial_p g^{(0)}(\xi,p)\,d\xi \\
&\simeq g^{(0)}(x,p) + \beta(p) \int_0^{\infty} u(\xi) g^{(0)}(\xi,p) \partial_{\xi} G_2^{(1)}(x,\xi;p)\,d\xi \\
&= g^{(0)}(x,p) + \beta(p) \frac{-g^{(0)}(x,p) e^{\tau_{\text{esc}}^{-1}\Phi(x)}}{D} \\
&\qquad \times \int_0^x u(\xi) e^{-\tau_{\text{esc}}^{-1}\Phi(\xi)}\,d\xi \\
&\quad + \frac{\tau_{\text{esc}}^{-1}\beta g^{(0)}(x,\xi) e^{\tau_{\text{esc}}^{-1}\Phi(x)}}{D^2} \\
&\qquad \times \int_{0}^{\infty}u(\xi)e^{-\tau_{\text{esc}}^{-1}\Phi(\xi)}d\xi \\
&\qquad \times \int_{\xi}^{\infty} e^{-\phi(\eta)/D}\,d\eta \int_0^{\min(x,\eta)} e^{\phi(s)/D}\,ds \\
&\simeq g^{(0)}(x,p) e^{-\beta[\phi(x)/D - \tau_{\text{esc}}^{-1}\Phi(x)]}
\end{aligned}
\label{eq50}
\end{equation}

where

\begin{equation}
g^{(0)}(x, p) = g_0 e^{-\phi(x)/D - \tau_{\text{esc}}^{-1} \Phi(x)}
\label{eq51}
\end{equation}

is a first-order estimate of the homogeneous solution to \(L_2\), and

\begin{equation}
\Phi(x) \equiv \int_0^x \frac{d\xi}{u(\xi)}
\label{eq52}
\end{equation}

\begin{equation}
\beta(p) = -\frac{p}{3} \frac{\partial \ln g_0}{\partial \ln p} = \frac{q}{3} - 1
\label{eq53}
\end{equation}

From

\begin{equation}
\begin{split}
\frac{du}{dx} &\propto - \int \frac{p\partial_x g(x, p) \, dp}{\sqrt{p^{2}+1}} \\
&\sim \int \frac{p \, dp}{\sqrt{p^2 + 1}} \frac{u g}{D} 
      \left( 1 + \beta \left( 1 - \left( \frac{v_{\text{esc}}}{u} \right)^2 \right) \right)
\end{split}
\label{eq54}
\end{equation}

the above calculation is applicable when \(v_{\text{esc}} < u\). Outside this region, higher-order corrections to \(G_2\) need to be considered. However, in any case, we can see that the presence of the escape term suppresses the high-energy contribution to \(\frac{du}{dx}\) before the escape suppresses the upstream distribution, that is, it suppresses the feedback effect.

A similar behavior can also be seen with finite boundaries. Setting an escape boundary at \(x = L\) upstream: \(g(\xi, p) = 0\). The corresponding Green's function is calculated as:

\begin{equation}
G(x, \xi; p) = G_1(x, \xi; p) F(\max(x, \xi), p)
\label{eq55}
\end{equation}

where the boundary correction factor is

\begin{equation}
F(x, p) = 1 - \frac{\int_0^x e^{\phi(t)/D} /D\, dt}{\int_0^L e^{\phi(s)/D}/D \, ds}
\label{eq56}
\end{equation}

which decreases as momentum increases. At very high energies (beyond the cutoff momentum), it suppresses the system's response almost linearly:

\begin{equation}
G(x, \xi; p \gg 1) \sim G_1(x, \xi; p) \left( 1 - \frac{x}{L} \right)
\label{eq57}
\end{equation}

In the same manner, we have:

\begin{equation}
g(x,p) = e^{-\phi(x,p)} g_{0}(p)F(x,p)+\int_{0^{+}}^{L}S[g](x',p)G(x,x',p)dx' 
\label{eq58}
\end{equation}

and $g_{0}$ satisfies:

\begin{equation}
\begin{aligned}
g_{0}(p)\Bigl[u_2&+\Bigl(\int_{0}^{L}\frac{e^{\phi(x,p)}dx}{D}\Bigr)^{-1}\Bigr] \\
&+\frac{p\Delta u}{3}\partial_{p}g_{0}(p) \\
&+\int_{0^{+}}^{L}S[x,p]F(x,p)dx = 0
\end{aligned}
\label{eq59}
\end{equation}

\begin{equation}
    \phi(x,p) = \int_{0}^{x}\frac{u(x')dx'}{D}
\label{eq60} 
\end{equation}

\begin{equation}
    S[g](x,p)=\frac{p}{3}\frac{du}{dx}\partial_{p}g(x,p)
    \label{eq61}
\end{equation}

Estimate from Eq.~\eqref{eq58}:

\begin{equation}
g^{(1)}(x, p) \simeq g_0(p) e^{-\phi(x) \left( 1 + \beta F(x, p) \right) / D} F(x, p)
\label{eq62}
\end{equation}

\begin{equation}
\frac{du}{dx} \propto \int \frac{p \, dp}{\sqrt{p^2 + 1}} \frac{u g}{D} \left( 1 + \beta F(x, p) \right)
\label{eq63}
\end{equation}

Thus, due to the escape boundary, the contribution of high-energy particles to the flow field feedback is also suppressed. In the region about one diffusion scale \(L_{\text{diff}} \sim D/u\) before the escape boundary \(L\), the contribution decays almost linearly. This suppression effect becomes important when \(\frac{L - L_{\text{diff}}}{L_{\text{diff}}} < 1\), roughly, for particles with momentum around the escape momentum before escape begins.

%
\end{document}